\documentclass[fleqn,usenatbib]{mnras}

\usepackage{newtxtext,newtxmath,hyperref}
\usepackage{newtxmath}
\usepackage[T1]{fontenc}

\DeclareRobustCommand{\VAN}[3]{#2}
\let\VANthebibliography\thebibliography
\def\thebibliography{\DeclareRobustCommand{\VAN}[3]{##3}\VANthebibliography}

\usepackage{graphicx}	
\usepackage{amsmath}	
\usepackage[english]{babel}
\usepackage{gensymb}
\usepackage{bm}

\title[The impact of beams on the bispectrum phase]{Simulations of primary beam effects on the cosmic bispectrum phase observed with the Hydrogen Epoch of Reionization Array}

\author[N.~Charles et al.]
{
\parbox{\textwidth}{
N.~Charles$^{1}$\thanks{ntsikelelo.charles@gmail.com}, G.~Bernardi$^{2,1,3}$, H.L.~Bester$^{3,1}$, O.M.~Smirnov$^{1,3}$, C.~Carilli$^{4}$, P.M.~Keller$^{5}$, N.~Kern$^{6}$, B.~Nikolic$^{5}$, N.~Thyagarajan$^{7,4}$, E.~de Lera Acedo$^{5,8}$, N.~Fagnoni$^{5}$, M.G.~Santos$^{9,3}$}
\vspace{0.4cm} \\
\parbox{\textwidth}{
$^{1}$Department of Physics and Electronics, Rhodes University, PO Box 94, Makhanda, 6140, South Africa\\
$^{2}$INAF-Istituto di Radio Astronomia, via Gobetti 101, 40129 Bologna, Italy\\
$^{3}$South African Radio Astronomy Observatory, Black River Park, 2 Fir Street, Observatory, Cape Town, 7925, South Africa\\
$^{4}$ National Radio Astronomy Observatory, 1003 Lopezville Road, Socorro, NM 87801, USA\\
$^{5}$Cavendish Astrophysics, University of Cambridge, Cambridge CB3 0HE, UK\\
$^{6}$Dept. of Physics and Kavli Institute for Astrophysics and Space Research, Massuchusetts Institute of Technology, Cambridge, MA, USA\\
$^{7}$Commonwealth Scientific and Industrial Research Organisation (CSIRO), Space $\&$ Astronomy, PO Box 1130, Bentley, WA 6102, Australia \\
$^{8}$Kavli Institute for Cosmology in Cambridge, University of Cambridge, Cambridge, UK\\
$^{9}$Department of Physics and Astronomy, University of Western Cape, Cape Town 7535, South Africa.
}
}

\pubyear{2020}

\begin{document}
\label{firstpage}
\pagerange{\pageref{firstpage}--\pageref{lastpage}}
\maketitle

\begin{abstract}
The 21~cm transition from neutral Hydrogen promises to be the best observational probe of the Epoch of Reionisation. The main difficulty in measuring the 21 cm signal is the presence of bright foregrounds that require very accurate interferometric calibration. Closure quantities may circumvent the calibration requirements but may be, however, 
affected by direction dependent effects, particularly antenna primary beam responses. This work investigates the impact of antenna primary beams affected by mutual coupling on the closure phase and its power spectrum. Our simulations show that primary beams affected by mutual coupling lead to a leakage of foreground power into the EoR window, which can be up to $\sim 10^4$ times higher than the case where no mutual coupling is considered. This leakage is, however, essentially confined at  $k < 0.3$~$h$~Mpc$^{-1}$ for triads that include 29~m baselines. The leakage magnitude is more pronounced when bright foregrounds appear in the antenna sidelobes, as expected. Finally, we find that triads that include mutual coupling beams different from each other have power spectra similar to triads that include the same type of mutual coupling beam, indicating that beam-to-beam variation within triads (or visibility pairs) is not the major source of foreground leakage in the EoR window.

\end{abstract}




\section{Introduction}
\label{sec:intro}

The detection of the redshifted 21~cm emission line from neutral Hydrogen during the Epoch of Reionisation (EoR) is one of the main goals of (current and upcoming) low frequency radio telescopes like the Low-Frequency Array \citep[LOFAR,][]{Van-Haarlem2013}, the Murchison Widefield Array \citep[MWA,][]{Tingay2013}, the Giant Metrewave Radio Telescope EoR experiment \citep[GMRT,][]{Paciga2013}, the Hydrogen Epoch of Reionisation Array \citep[HERA,][]{DeBoer2017}, and the Square Kilometre Array \citep[SKA,][]{Koopmans2015}. The EoR is one of the least known areas of cosmology, from an observational point of view. Advancing our understanding of reionisation will enable us to understand how the first galaxies formed and, ultimately, improve constraints on cosmological parameters \citep[e.g.,][]{Furlanetto2006,Mesinger2016,Park2019,Zara2021}.

The hyperfine transition from neutral Hydrogen (21~cm emission) is one of the most promising probes of structure formation, imprinted in the intergalactic medium. Measurements of the 21~cm signal are challenged by the presence of foreground emission from the Galaxy and extra-galactic sources, which are orders of magnitude brighter \citep[e.g.,][]{Santos2005,Bernardi2009,Ali2015}. Foregrounds are spectrally smooth, unlike the 21~cm emission line which fluctuates rapidly \citep[e.g.,][]{Santos2005}. If foreground spectral and spatial characteristics are known, they can be subtracted to isolate the 21~cm emission. The process generally begins with the subtraction of bright, compact sources, After bright source subtraction, the sky brightness is dominated by the diffuse foreground emission \citep[i.e.,][]{Bernardi2010,Pober2013,Dillon2014}, which can be subtracted leveraging, again, on the foreground spectral smoothness \citep[e.g.,][]{Mertens2018,Ghosh2020,Kern2021}. In practice, though, smooth spectrum foregrounds are corrupted by instrumental effects. The calibration process attempts to correct for these corruptions. High accuracy in calibration is therefore required in order not to compromise the foreground spectral smoothness \citep[e.g.,][]{Wang2013,Chapman2015,Sims2016,Datta2017,Kern2021}. Additionally, some of the brightest sources can have complicated, extended morphologies: failing to model and subtract them accurately can leave residual foreground contamination that may prevent the 21~cm detection. Over-fitting diffuse emission may equally lead to 21~cm signal loss \citep[e.g.,][]{Wang2013,Cheng2018}. 

The calibration process makes use of a sky model to correct for instrumental effects \citep{Smirnov2011}. Sky models are built from catalogues of compact sources with known properties, and often cover an area larger than the field of view of the observation \citep{Yatawatta2013,Pober2016}. The sky model ideally should contain the entire sky emission but, inevitably, remains incomplete due to the limited angular resolution and depth of existing catalogues \citep{Grobler2014,Wijnholds2016,Trott2016,Procopio2017,Barry2021}. Calibration errors due to incomplete sky models lead to leakage of foreground power into the EoR window \citep{Barry2016,Ewall-Wice2017}.
 
The need for highly accurate calibration required for foreground subtraction has led to alternate methods, known as \textit{foreground avoidance methods}. As the name suggests, the idea is to avoid the foreground emission rather than subtracting it \citep[e.g.,][]{Parsons2012,Thyagarajan2013}. The \textit{delay spectrum} is one such method; it makes use of interferometric delays to isolate the power spectrum of the 21~cm emission \citep{Parsons2012}. Due to the spectral nature of the 21~cm signal, its power spectrum appears at all $k$ modes, whereas the foreground emission is limited to a wedge like region in $k-$space \citep{Datta2010,Parsons2012,Vedantham2012,Trott2012,Thyagarajan2013,Hazelton2013,Pober2013, Liu2016,Morales2019}. Foreground avoidance methods remain, however, prone to calibration errors.

Yet another alternative method to detect the 21~cm signal was proposed by \cite{Thyagarajan2018} and takes advantage of closure quantities. The use of closure phases mitigates calibration requirements as closure quantities are independent (to first order) of antenna based corruptions. In terms of foregronud separation,  \citet{Thyagarajan2020} showed that the dynamic range required to detect the 21~cm signal is similar to the standard power spectrum approach \citep[e.g.,][]{Parsons2012,Abdurashidova2021}. For a massively redundant array like HERA \citep{Dillon2015,DeBoer2017} closure quantities may, therefore, represent an appealing alternative to the mainstream power spectrum analysis.

Closure phase quantities, however, are affected by direction-dependent effects such as varying antenna primary beams due to mutual coupling induced by a very compact configuration \citep[e.g.,][]{Fagnoni2021,Josaitis2021}. Variations of primary beams across the array invalidates the assumption of redundancy (i.e., that baselines with the same length and orientation measure exactly the same signal from the sky) which is the core of the HERA calibration strategy \citep{Dillon2020}. Several authors have empirically simulated the impact that deviations from redundancy have on the calibration and proposed possible mitigation schemes \citep{Ewall-Wice2017,Joseph2018,Orosz2019,Choudhuri2021}. It has been established that variations from redundancy due to enhanced structure in primary beams couples foreground structure observed through sidelobes into spectral structure in calibration solutions \citep{Orosz2019,Kern2020,Choudhuri2021}.

The analysis of HERA data using closure phase also showed some evidence of deviation from redundancy \citep{Carilli2018}, in particular the presence of a baseline-dependent systematic effect appearing at $k_\parallel\sim 0.5$~$h$~Mpc$^{-1}$ \citep{Thyagarajan2020}. At higher $k$ modes, however, the closure phase analysis of $\sim 2$~h of HERA observations shows no evidence of systematic effects, suggesting that longer integrations may reduce the thermal noise \citep{Thyagarajan2020}.

In this paper we simulate the impact that different primary beams have on closure phase in the case of HERA observations, specifically investigating the case when two beams are different within a baseline pair. We quantify the effect that such deviations from redundancies have on the power spectrum of the bispectrum phase of foreground emission and the impact on the EoR window.

The paper is organized as follows: Section~\ref{sec:formalism} summarizes the closure phase formalism, Section~\ref{sec:simulations} describes our simulations, Section~\ref{sec:results} presents the simulated closure spectra and power spectra and we conclude in Section~\ref{sec:conclusions}.

\section{Closure phase formalism} 
\label{sec:formalism}

The simplest radio interferometer is the two-element interferometer, where signals measured from a pair of antennas $(p,q)$ are cross-multiplied and averaged in time. This operation is known as \textit{correlation} and leads to the fundamental quantity measured in radio interferometry, the \textit{visibility function} $V_{pq}$. The \textit{Van Cittert-Zernike theorem} states that the correlation of signals from the $(p,q)$ pair is related to the sky brightness $I(\bm{s},\nu)$ by a Fourier transform like relation:
\begin{equation}
V_{pq}(\nu) = \iint I(\bm{s},\nu) \, \exp{(-2\pi \imath \frac{\nu}{c} \bm{b}_{pq} \cdot \bm{s})} \,\frac{dldm}{n(\bm{s})}, 
\label{eq:zernike_theorem}
\end{equation}
where $\bm{b}_{pq}$ is the baseline vector connecting antenna $p$ and $q$, $\bm{s} = [l, m, n]^T$ is a unit vector (so that $n = \sqrt{1 - l^2 - m^2}$ with $(l, m, n)$ the direction cosines of $\bm{s}$) representing a direction on the celestial sphere, $\nu$ is the observing frequency and $c$ the speed of light.

In real observations, signals are corrupted by the antenna response. Corruptions are modeled using antenna-based gain terms \citep[the so called measurement equation][]{Smirnov2011} that can include the antenna primary beam pattern $E$. The antenna primary beam depends upon the observing direction, frequency and time - the latter normally due to the rotation of the sky with respect to the feed orientation.
In this paper we investigate the response of a single polarisation feed, for which the measurement equation takes the following form:
\begin{equation}
V_{pq}(\nu) = \iint J_p(\bm{s},\nu) \, I(\bm{s},\nu) \, J^*_q(\bm{s}, \nu) \, K_{pq}(\bm{s},\nu) \,\frac{dldm}{n(\bm{s})}, 
\label{eq:final_discrete_rime}
\end{equation}
where $J_p=G_p E_p$ is a general antenna gain that can include a direction independent contribution $G_p$ as well as the direction dependent antenna primary beam pattern $E_p$ and we have introduced the abbreviated notation $K_{pq}(\bm{s},\nu)$ representing the exponential term introduced in \eqref{eq:zernike_theorem}. It is normally assumed that the primary beam is the same for the two receiving elements, i.e. $E_p = E_q$, but here we explicitly explore beams that are different from each other, i.e. $E_p \ne E_q$.

In this work we assume that the sky can always be modelled as a collection of point sources so that equation~\ref{eq:final_discrete_rime} can be discretised as:
\begin{equation}
    {V}_{pq\nu} = \sum_s {J}_{ps\nu} \, I_{s\nu} \, {J}^*_{qs\nu}  \, K_{pqs\nu} = \sum_s {J}_{ps\nu} \, X_{pqs\nu} \, {J}^*_{qs\nu},
    \label{eq:beam_corruption_of_models}
\end{equation}
where $s$ labels sources so that $I_{s\nu}$ is the point source flux density $I(\bm{s},\nu)$ in the direction of source $s$ at frequency $\nu$, for example. Here we have also introduced the source coherency $X_{pqs\nu}$ corresponding to the visibilities of a specific source. As shown below, closure phases are independent of direction independent antenna based gains. Thus we use equation~\ref{eq:beam_corruption_of_models} with $J_p = E_p$ to apply differing primary beams to simulated sky models. Equation~\ref{eq:beam_corruption_of_models} is implemented efficiently using the \textit{codex-africanus} package \citep{Perkins2021} and written to a measurement set format using \textit{dask-ms} \citep[see,][]{daskms}.

At this point we introduce the visibility bispectrum $C_{pqr}$, defined as the triple product of three visibilities from baselines $(pq, qr, rp)$:
\begin{equation}
    C_{pqr} = V_{pq} V_{qr} V_{rp},
\end{equation}
where indices $p,q,r$ are antenna labels. Assuming that antenna gains consist of purely direction independent gains, we can rewrite equation \eqref{eq:beam_corruption_of_models} as
\begin{equation}
    V_{pq\nu} = G_{p\nu} X_{pq\nu} G^*_{q\nu},
\end{equation}
where the individual source coherencies have been combined into a single model coherency term i.e. $X_{pq\nu} = \sum_s X_{pqs\nu}$.
The bispectrum then becomes:
\begin{eqnarray}
    C_{pqr} = |G_p|^2 \, |G_q|^2 \, |G_r|^2 \, X_{pq} \, X_{qr} \, X_{rp},  
    \label{eq:bispectrum}
\end{eqnarray}
as the antenna-gain phases cancel out. Splitting the model coherencies into amplitudes and phases
\begin{equation}
    X_pq = |X_{pq}| \exp (\imath \phi_{pq}),
\end{equation}
we see that the phase of the bispectrum (also known as the closure phase):
\begin{equation}
    \phi_\triangledown = \phi_{pq} + \phi_{qr} + \phi_{rp},
\end{equation}
is independent of antenna based direction independent gains. Here we use $\phi_\triangledown$ to denote the bispectrum phase of a closed triad. It is worth noticing that closure quantities are not immune from spectral structure imparted on the sky signal by the instrument - as it will appear from simulations carried out in this work.
 
Closure phases will contain contributions from both foregrounds and the cosmological signal of interest. \citet{Thyagarajan2018} showed that it is possible to separate the two contributions by leveraging the different frequency behaviour of the foreground and 21~cm signal closure spectra. In analogy with the delay spectrum approach \citep{Parsons2012,Pober2013}, they suggested to form a power spectrum $P_\triangledown$ by taking the Fourier transform along the frequency axis of the complex bispectrum phase \citep{Thyagarajan2020}:

\begin{equation}
    P_\triangledown(k_{||})= |\Tilde{\Psi}_\triangledown|^2 \left( \frac{\lambda^2}{2 k_B} \right)^2 \left( \frac{D_c^2 \, \Delta D_c}{B_{\rm eff}} \right) \left( \frac{1}{\Omega \, B_{\rm eff}} \right),
\end{equation}
where  $\Delta D = \Delta D(z)$ is the comoving depth along the line of sight corresponding to an effective bandwidth $B_{\rm{eff}}$, $\Omega=A_e/\lambda^2$ where $A_e$ is the effective aperture area and: 
\begin{equation}
    \Tilde{\Psi}_\triangledown=\Tilde{W}(\tau)*\Tilde{\Xi}_\triangledown(\tau)* V_{\rm eff}*\delta(\tau),
\end{equation}
where $*$ denotes convolution, $\Tilde{W}$ and $\Tilde{\Xi}_\triangledown$ are the delay transforms of the window function $W$ and the complex closure phase $\Xi_\triangledown$ respectively:
\begin{equation}
    \Xi_\triangledown (\nu)=e^{\imath \phi_\triangledown(\nu)}.
\end{equation}
In this paper we used a Blackman Harris window function \citep[e.g.,][]{Parsons2012,Thyagarajan2013} an effective bandwidth $B_{\rm eff} = 9.77$~MHz, centred at 175~MHz ($z = 7.1$) with a 97.66~kHz channel width i.e. the same observing setup as HERA. We also simulate a single snapshot observation.
The normalization factor $V_{\rm eff}$ is defined as \citep{Thyagarajan2020a}:
\begin{equation}
    (V_{\rm eff})^{-2} = \sum_{b=1}^3 |V_{b}'|^{-2},
\end{equation}
where
\begin{equation}
    V_{b}' = \frac{\int W(\nu) \, V_{b}(\nu) \, d\nu}{\int W(\nu) \, d\nu},
\end{equation}
and $b$ denotes baselines in a triad.
The closure spectrum is a dimensionless quantity and the normalisation factor calibrates the power spectra of fields with different brightness distribution on the same scale. We note that the power spectrum of the bispectrum phase is not directly comparable with the standard power spectrum \citep[e.g.,][]{Abdurashidova2021}, even if they share the same units.

In this paper we construct different power spectra of the bispectrum phase by simulating visibilities with beams that are different for each receptor using equation~\ref{eq:beam_corruption_of_models}.

\section{Bispectrum phase simulations}
\label{sec:simulations}

\subsection{Beam models}
  
\begin{figure}
    \centering
    \begin{tabular}{c}
        \includegraphics[scale=0.3]{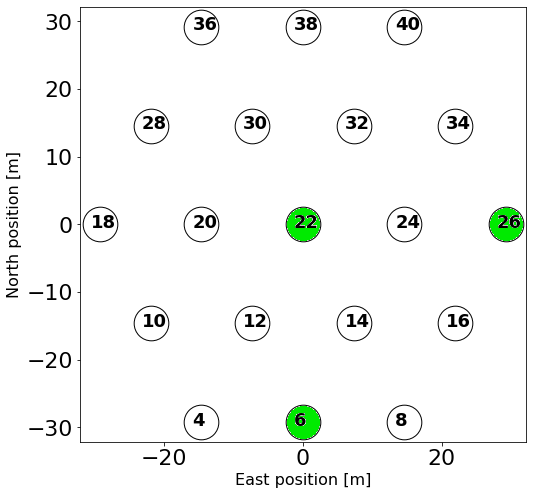} 
    \end{tabular}
    \caption{Simulated~HERA 19 array layout. In this work, we used the simulated primary beam pattern corresponding to label 22, for a central dish and beam patterns corresponding to two edge antennas, i.e. label 26 and 6. We only make use of $Y$ polarization (North-South) primary beam patterns.
    }
    
    
    \label{fig:Nicholas_mutual_coupling_beam_array_layout}
\end{figure}
\begin{figure}
    \centering
    \begin{tabular}{c}
        \includegraphics[scale=0.5]{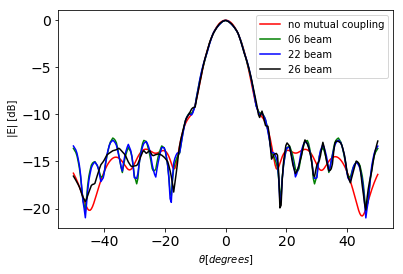} 
    \end{tabular}
    \caption{Cut through the HERA primary beam models at $\nu=175$~MHz for the $Y$ polarisation. Whereas the main lobe structure remains essentially the same for all beams, sidelobes have a more prominent structure for beams with mutual coupling and appear asymmetric in the case of antenna 26.}
    \label{fig:Nicholas_mutual_coupling_beam}
\end{figure}
HERA antennas consist of a dipole feed suspended above a parabolic dish with a diameter of 14~m. The dish structure was initially designed by paying specific attention to its spectral response, i.e. keeping the dish reflections and passband sufficiently smooth so that the EoR window would be preserved at $k_\parallel>0.2$~$h$~Mpc$^{-1}$ modes \citep{Ewall-Wice2016,Thyagarajan2016}. Further electromagnetic simulations \citep{Fagnoni2021} generated a primary beam model that is routinely used in the analysis and simulations of HERA observations  \citep[e.g.,][]{Martinot2018,Kern2020}. Due to compactness of the array, however, effects such as cross coupling amongst antennas, i.e. mutual coupling,  cause deviations to the ideal antenna model. 

\citet{Fagnoni2021} also carried out simulations of HERA dishes which included the receiving system and the effects of mutual coupling for the two polarisations, i.e. $X$ and $Y$, for a redundant, compact array layout that included 19, hexagonally-packed HERA dishes that were the first HERA installment \citep[][Figure~\ref{fig:Nicholas_mutual_coupling_beam_array_layout}]{Kohn2019}. Due to the interaction with many more dishes, mutual coupling may be subtly different for the full HERA array compared to the models employed in this work \citep{Dillon2015}. However, it would be surprising if subtle differences in the beam patterns significantly alter the derived power spectra. Thus we believe that our results should be qualitatively correct and should also hold for the full HERA array. 

\citet{Fagnoni2021} showed that mutual coupling  introduces extra sidelobe ripples (Figure~\ref{fig:Nicholas_mutual_coupling_beam}) and increases the sidelobe level by $2-4$~dB. Figure~23 in \citet{Fagnoni2021} shows that the gain at zenith  also varies as a function of frequency, up to $\sim 0.3$~dB with respect to the ideal beam and for different antenna positions within the array. The beam value at zenith oscillates with a periodicity of about 20~MHz, which corresponds to reflections occurring at 15~m path length, approximately the distance between the centre of two dishes. These effects lead to further deviations from the smooth ideal beam response. 

Lastly, antennas experience a varying degree of mutual coupling and, as a consequence, an antenna at the edge array has an asymmetric primary beam pattern since one side of antenna experiences more mutual coupling, i.e. side facing other dishes, than the other side where there are no dishes (Figure~\ref{fig:Nicholas_mutual_coupling_beam}, antenna 26).

\subsection{Simple sky models}

\begin{figure}
    \centering
    \begin{tabular}{cc}
           \includegraphics[scale=0.5]{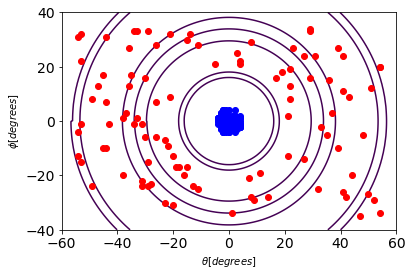} \\ 
    \end{tabular}
    \caption{Position of sources in simulated sky models, the blue dots mark point source positions within the beam main lobe and the red dots mark point source positions in the sidelobe region. Black lines mark nulls of the HERA ideal beam model at $\nu=175$~MHz. }
    \label{fig:random_artificial_sky_models}
\end{figure}
\begin{figure*}
    \centering
    \begin{tabular}{cc}
       \includegraphics[scale=0.55]{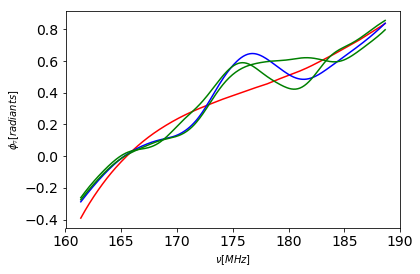} &            \includegraphics[scale=0.55]{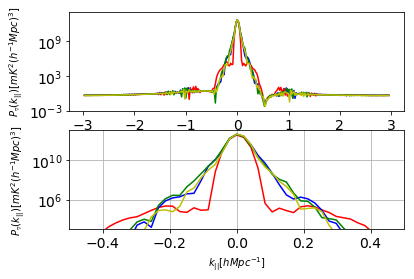}
    \\
         \includegraphics[scale=0.55]{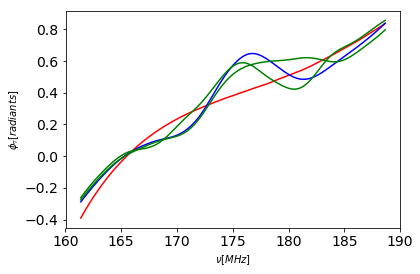} &            \includegraphics[scale=0.55]{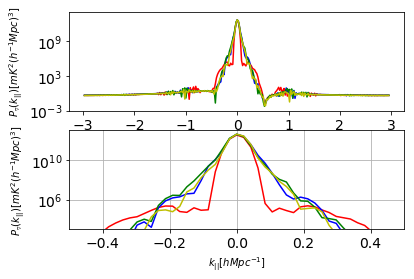}
        \\ 
         \includegraphics[scale=0.55]{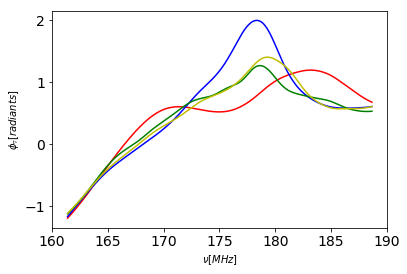}    &
         \includegraphics[scale=0.55]{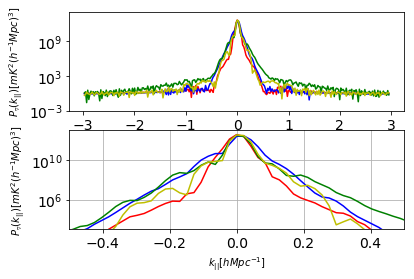} 
          
           \\
         \includegraphics[scale=0.55]{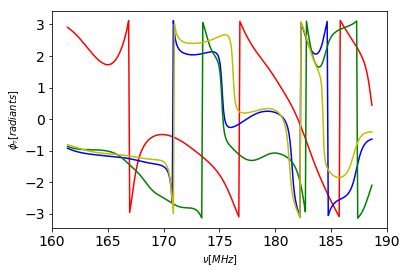} &
         \includegraphics[scale=0.55]{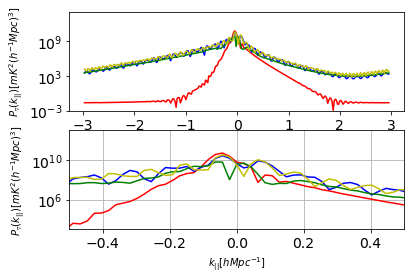}
    \end{tabular}
    \caption{Simulated closure spectra (left column) and power spectra of the bispectrum phase (right column) corresponding to sky models A0 (top row), A1 (second row), A2 (third row) and A3 (fourth row) - see text for details. The following triads are shown: $\triangledown_{HHH}$ (red), $\triangledown_{CCC}$ (blue), $\triangledown_{CCE}$ (yellow) and  $\triangledown_{CEE}$ (green). Bottom panels on the right column are zoom into the corresponding upper panels.}
    \label{fig:mutual_artificial_closure_spectrum_differing}
\end{figure*}
\begin{figure*}
    \centering
    \begin{tabular}{cc}
        \includegraphics[scale=0.55]{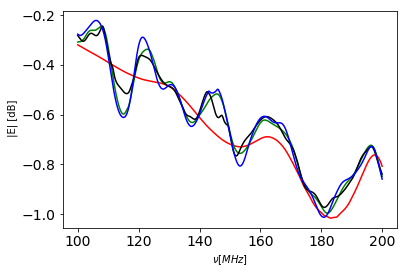} & \includegraphics[scale=0.55]{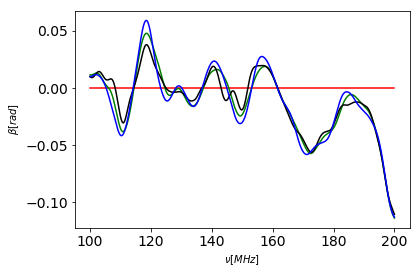}
        \\
          \includegraphics[scale=0.55]{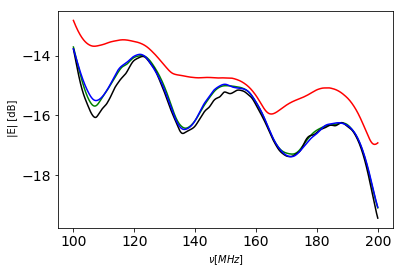} & \includegraphics[scale=0.55]{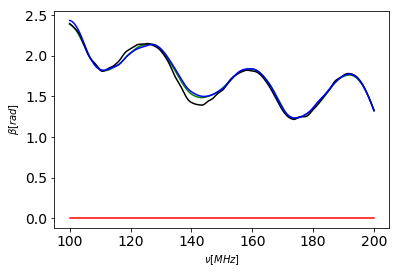}
    \end{tabular}
    \caption{Top row: the left panel shows beam response [dB]  averaged across all source locations for the sky model with sources only in the main lobe (blue dots in Figure~\ref{fig:random_artificial_sky_models}) as a function of frequency for $E_{06}$ (green), $E_{22}$ (blue), $E_{26}$ (black) and $E_{H}$ (red). The right panel corresponding beam phase as a function of frequency. Second row: same, but for sources in the sidelobe region (red dots in Figure~\ref{fig:random_artificial_sky_models}). }
  \label{fig:average_line_of_sigth_and_complex_phase}
\end{figure*}

We begin with simulating simplified sky models in order to demonstrate some basic properties of closure spectra and mutual coupling beams. We generate sky models where we randomly place 100 point sources in the main lobe and another 100 across the beam sidelobes (Figure~\ref{fig:random_artificial_sky_models}). All sources on the main lobe of the primary beam have a 1~Jy flux density at 150~MHz. We vary the flux density of the sources on sidelobes to create four different sky models, which we name A0, A1, A2 and A3 respectively. These sky models are meant to illustrate fields with faint and bright emission in the sidelobe area. Flux densities of sources corresponding to the different models are shown in Table~\ref{tab:simplistic_sky_models}. All sources have a spectral index $\alpha=0.7$, where   $\alpha$ is defined such that flux density $S$ of a source at frequency $\nu$ is given by:

\begin{equation}
    S(\nu) =S_0 \bigg (\frac{\nu}{\nu_0} \bigg)^{-\alpha},
\end{equation}
and $S_0$ is the flux density at some reference frequency $\nu_0$.
\begin{table}
    \centering
    \begin{tabular}{c|c|c}
    \hline
    Sky model ID    &   Flux density of sources &   Flux density of sources \\
                    &   in main lobe (Jy)       &   in the sidelobe region (Jy) \\ \hline
    
    A0  &   1       &   0 \\
    A1  &   1       &   0.01\\
    A2  &   1       &   1\\
    A3  &   0.01    &   1\\
    \end{tabular}
    \caption{Characteristics of the simple sky models used in simulations (Figure~\ref{fig:random_artificial_sky_models}).} 
    \label{tab:simplistic_sky_models}
\end{table}

We use beams from dishes 22, 26 and 6 (Figure~\ref{fig:Nicholas_mutual_coupling_beam_array_layout}) and simulate the effect of mutual coupling of an antenna placed at the centre of the array and two at the edge, respectively. Hereon, we denote primary beams from dish 22, 26, 6 and the HERA ideal beam, i.e. beam with no mutual coupling, as $E_{22}$, $E_{26}$, $E_{06}$ and $E_H$ respectively. We combine different primary beams to simulate four types of 29~m equilateral triads: (1) a triad at the centre of the array, with only $E_{22}$ beams ($\triangledown_{CCC}$); (2) one at the edge of the array with one centre beam $E_{22}$ and two different edge beams $E_{26}$ and $E_{06}$ ($\triangledown_{CEE}$); (3) a second triad at the edge of the array with two $E_{22}$ beams and one $E_{26}$ beam ($\triangledown_{ECC}$) and, finally, (4) a triad unaffected by mutual coupling with just $E_H$ beams ($\triangledown_{HHH}$).

We simulate noiseless visibilities. We acknowledge that, in practice, each dish has a unique beam as the mutual coupling varies across the array and the primary beam corresponding to dish 8 would be different than the primary beam for dish 6, for example. However, this approximation is acceptable for the scope of our investigation.

Figure~\ref{fig:mutual_artificial_closure_spectrum_differing} shows the corresponding closure spectra and power spectra of the bispectrum phase. We first consider sky models with faint or no sources in the sidelobe region, i.e., A0 and A1. Closure spectra of both models are essentially identical, as they are dominated by sources within the primary beam main lobe. Main lobes have a very similar spectral structure for all the beams, yielding very similar closure and power spectra. Among the triads, the one that includes only the ideal beam has the smoothest frequency behaviour - as expected. Similarly, the power spectrum of the $\triangledown_{HHH}$ triad has a very distinct behaviour: the power is concentrated at small $k_\parallel$ modes as it is expected for smooth spectrum foregrounds \citep{Thyagarajan2018} and falls already by $\sim 10^8$ times at $k_\parallel \sim0.1$~$h$~Mpc$^{-1}$.
Power spectra of triads with mutual coupling beams, conversely, show up to $\sim 10^4$~mK$^2$~($h^{-1}$~Mpc$)^{3}$ higher power starting already at $k_\parallel \sim0.1$~$h$~Mpc$^{-1}$. This is indeed indicative of excess spectral structure in the closure spectra of triads with mutual coupling, likely arising from the gain variation at zenith and the different phases for different beams.

As we increase the brightness of the sources in the sidelobe region, i.e. model A2, closure spectra from all triads show extra frequency structure compared to model A0 and A1, due to sidelobe ripples. This results in an excess power up $\sim 10^4$~mK$^2$($h^{-1}$~Mpc$)^{3}$ for triads with mutual coupling, notably at large $k_\parallel$ values, $0.5 < |k_\parallel| < 2$~$h$~Mpc$^{-1}$. It is worth noting that the leakage is  more pronounced for triads with different primary beam patterns i.e. the edge triads.

When we decrease the brightness of the sources in the main lobe, i.e. sky model A3, the leakage is much worse, with an excess power between $10^4$ and $10^8$~mK$^2$~($h^{-1}$~Mpc$)^{3}$ at $|k_\parallel|>0.1$~$h$~Mpc$^{-1}$. We also note that in sky model A3, the center triad, where we have no beam variation, shows an excess power leakage comparable to edge triads. This shows that a large portion power leakage in A3 is actually caused by ripples on the sidelobes of mutual coupling beams (see Figure~\ref{fig:Nicholas_mutual_coupling_beam}).

In summary, our simulations show that, in the presence of bright emissions on the sidelobes, we may expect the power to leak at high $k_\parallel$ values, and the power leakage increases with brightness of sources on the sidelobes. Indeed, previous work by \citet{Choudhuri2021} shows similar results, as well as analysis of HERA data \citep[e.g.,][]{Kern2020,Dillon2020}. In addition, our simulations also show that the presence of bright sources on the main lobe mitigates the power leakage by dominating the overall closure spectra. In the case of extremely bright sources on sidelobes, we may expect the sidelobe ripples from mutual coupling beams to contribute a large fraction of the power leakage observed.

\subsection{Simulations with realistic sky models}
\begin{figure}
    \centering
    \begin{tabular}{cc}
         \includegraphics[scale=0.3]{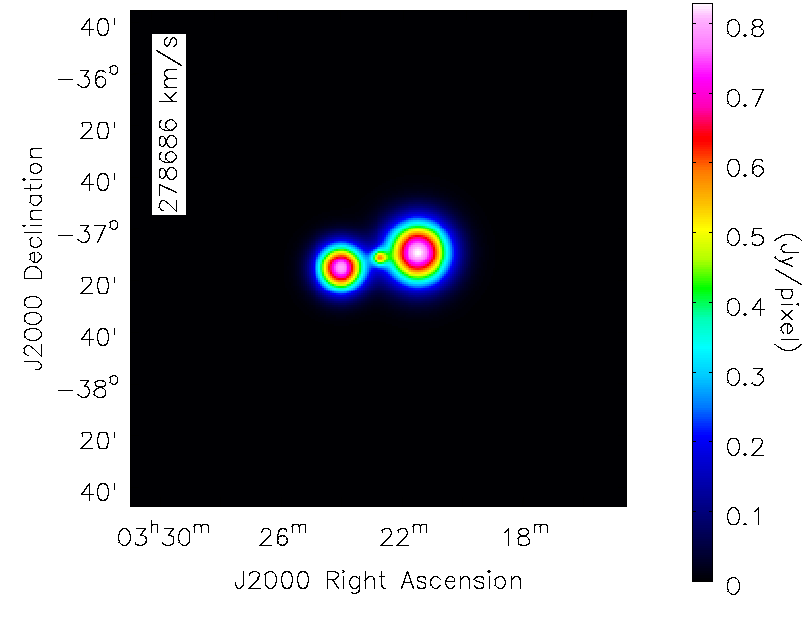}  &  
    \end{tabular}
    \caption{Fornax~A model image at $\nu=100$ MHz \citep[from][]{McKinley2015}. Units are Jy~pixel$^{-1}$ with a pixel size $=0.75$~$\mathrm{arcmin}$.}
    \label{fig:fornax_A_image_and_own_haslam_map}
\end{figure}
\begin{figure*}
    \centering
    \begin{tabular}{cc}
         \includegraphics[scale=0.45]{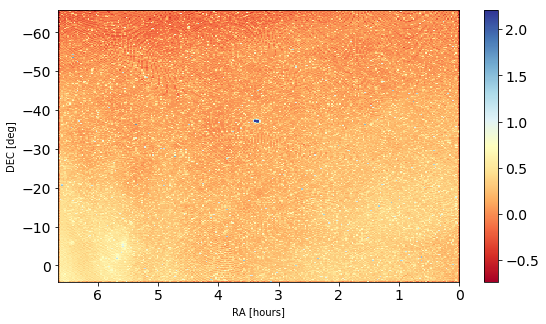} &
          \includegraphics[scale=0.45]{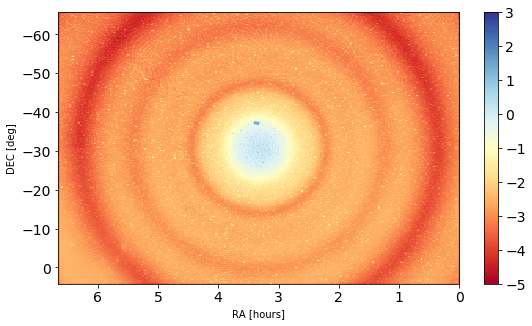}
         \\
         \includegraphics[scale=0.45]{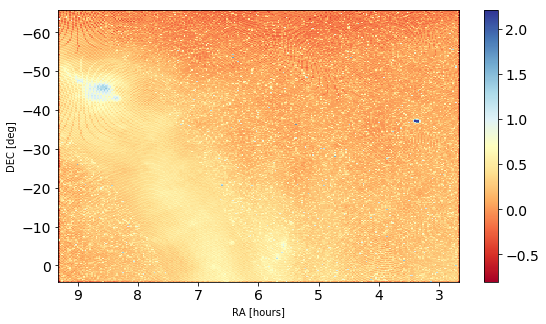} &
          \includegraphics[scale=0.45]{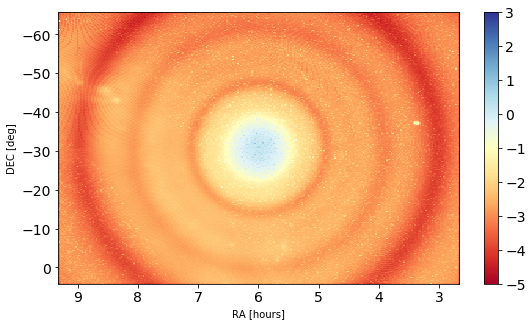}
         \\
         \includegraphics[scale=0.45]{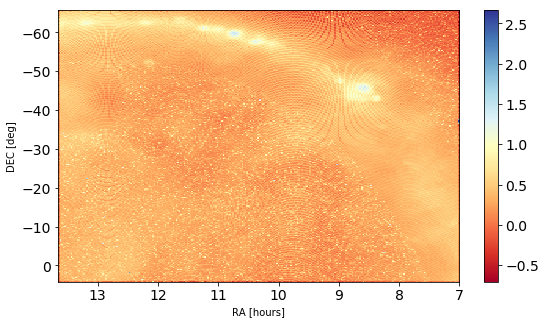}&
          \includegraphics[scale=0.45]{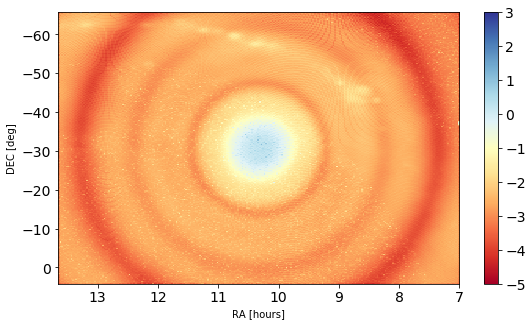} 
    \end{tabular}
    \caption{Left column shows sky model images of simulated fields, field~1 (first row), field~2 (second row) and field~3 (third row). Right columns shows the corresponding apparent sky model, i.e. after the primary beam, $E_{H}$ is applied. Units are $\log_{10}{|I(\mathrm{Jy~pixel}^{-1})|}$ .}
    \label{fig:full_sky_models}
\end{figure*}
After we have treated simplified sky models, we simulate three realistic zenith-pointed observations centred at right ascension $\alpha = (3^{\rm h} \, 20^{\rm m} \, 6.7^{\rm s})$, $(5^{\rm h} \, 20^{\rm m} \, 6.7^{\rm s})$ and ($10^{\rm h} \, 20^{\rm m} \, 6.7^{\rm s})$, that we label field~1, field~2 and field~3 respectively. They are located within the stripe observed by HERA \citep{Abdurashidova2021}. As mentioned in Section~\ref{sec:formalism}, we only simulate single-snapshot observations.
We include three sky model components for each pointing:
\begin{itemize}
    \item all point sources brighter than 200~mJy at 151~MHz and within a $100^\circ \times 70 ^\circ$ region around the centre of each pointing, taken from the GLEAM catalogue \citep{Hurley-Walker2017};
    \item Fornax~A - which is not included in the GLEAM catalogue. The source is modeled as a core and two lobes, based on observations at 174~MHz \citep[][ Figure~\ref{fig:fornax_A_image_and_own_haslam_map}]{McKinley2015}. The core is modelled with a circular Gaussian with a $5'$ axis, a $\alpha = 1$ spectral index and a 12~Jy flux density at 154~MHz. The West lobe is modelled with a circular Gaussian with a $20'$ axis, a $\alpha = 0.77$ spectral index and a 260~Jy flux density at 154~MHz. The East lobe is modelled with a circular Gaussian with a $15'$, a $\alpha = 0.77$ spectral index and a 480~Jy flux density at 154~MHz. Visibilities are generated using equation~\ref{eq:beam_corruption_of_models}, treating each image pixel as a point source;  
    \item an all-sky map of Galactic diffuse emission at 408~MHz \citep{Remazeilles2015} with a $56'$ resolution. The map (in the Healpix format) was extrapolated to 150~MHz using a spatially constant spectral index $\alpha = 0.7$. Like the Fornax~A case, each Healpix pixel is treated as a point source in equation~\ref{eq:beam_corruption_of_models}.
\end{itemize}

\section{Results}
\label{sec:results}
\begin{figure*}
    \centering
    \begin{tabular}{cc}
    \includegraphics[scale=0.6]{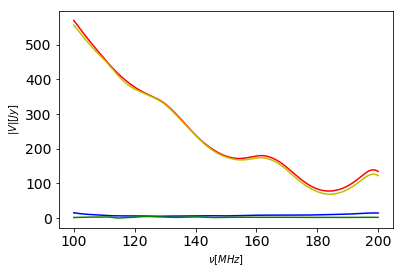} & \includegraphics[scale=0.6]{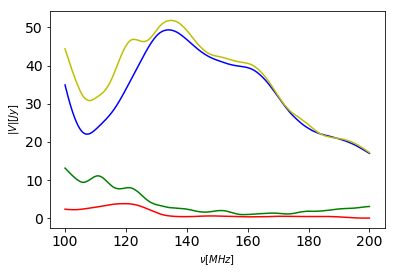}
    \\
    \includegraphics[scale=0.6]{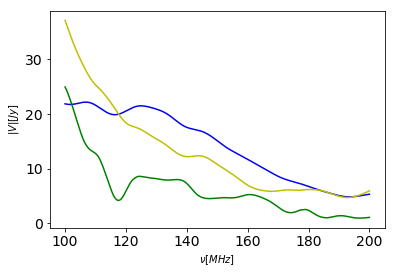}&
    \includegraphics[scale=0.6]{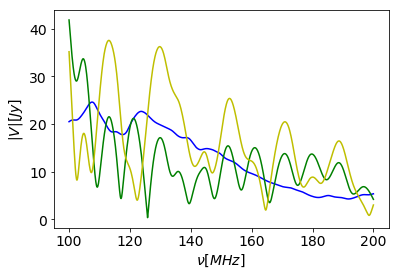}
    \end{tabular}
    \caption{Simulated visibility spectra corresponding to sky models for field~1 (top left panel), field~2 (top right panel) and field~3 (bottom left) for triad $\triangledown_{HHH}$. Colours indicate sky model components: Fornax~A (red),  GLEAM sources (blue), diffuse emission (green) and full sky model (Fornax A + GLEAM sources + diffuse emission; yellow). Note that Fornax~A is not included in field~3. The bottom right panel shows simulated visibility spectra for triad $\triangledown_{CCC}$ corresponding to the sky model for field~3.}
    \label{fig:visibilty_spectra_components}
\end{figure*}
\begin{figure*}
    \centering
    \begin{tabular}{cc}
         \includegraphics[scale=0.6]{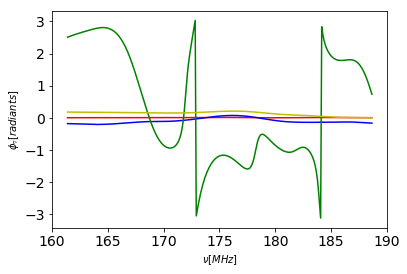} &
          \includegraphics[scale=0.6]{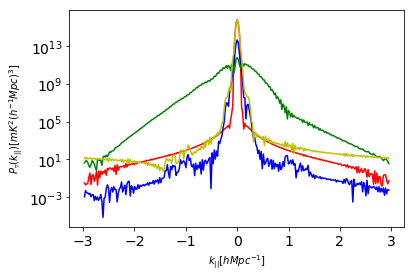}
         \\
         \includegraphics[scale=0.6]{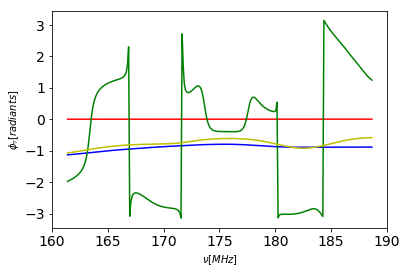} &
          \includegraphics[scale=0.6]{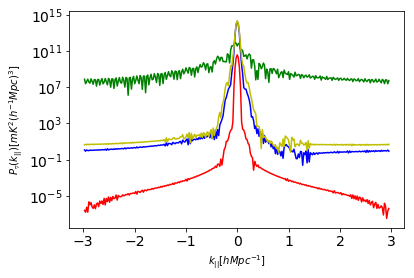}
          \\
         \includegraphics[scale=0.6]{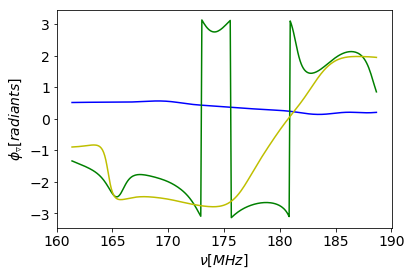} &
         \includegraphics[scale=0.6]{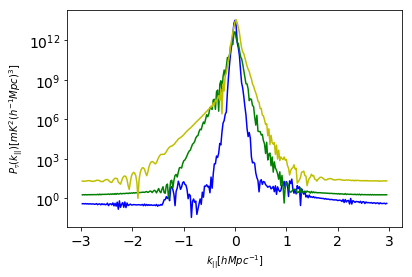}
         \end{tabular}
      \caption{Simulated closure spectra (left column) and power spectra (right column) corresponding to sky models field~1 (first row), field~2 (second row) and field~3 (third row) from triad $\triangledown_{CCC}$. Colours indicate sky model components:  Fornax~A (red),  GLEAM sources (blue), diffuse emission (green) and full sky model (Fornax~A, GLEAM sources and diffuse emission; yellow). Note that Fornax~A is not included in field~3.}
      \label{fig:GLEAM_1h_200mJy_closure_spectrum_average_component}
\end{figure*}
\begin{figure*}
    \centering
    \begin{tabular}{cc}
         \includegraphics[scale=0.6]{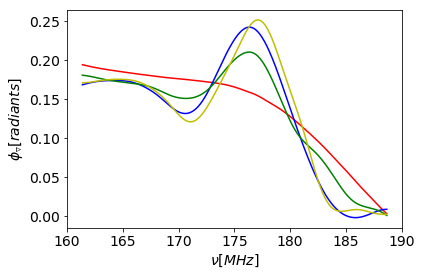} &
         \includegraphics[scale=0.6]{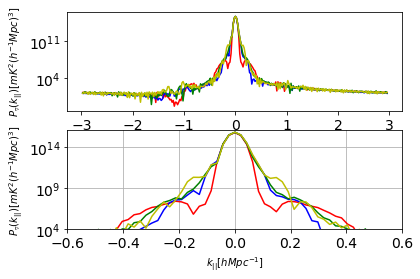}
         \\
         \includegraphics[scale=0.6]{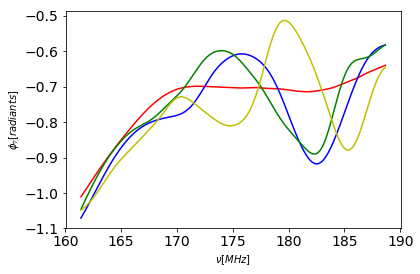} &
         \includegraphics[scale=0.6]{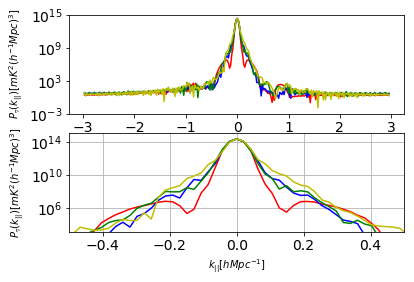}
         \\
          \includegraphics[scale=0.6]{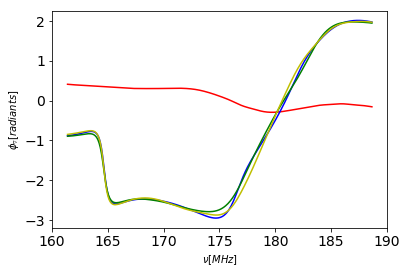} &
         \includegraphics[scale=0.6]{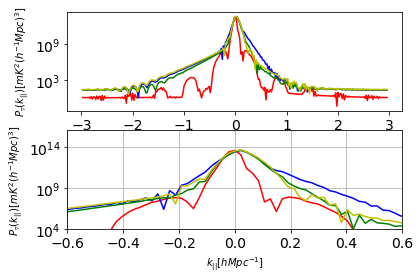}
         \end{tabular}
      \caption{Left column: closure spectra corresponding to sky models field 1 (first row), field 2 (second row) and field 3 (third row). The colours indicate here, the closure spectrum from different simulated triads: $\triangledown_{HHH}$ (red), $\triangledown_{CCC}$ (blue), $\triangledown_{CEE}$ (green) and $\triangledown_{ECC}$ (yellow). Right column: corresponding power spectra.}
      \label{fig:GLEAM_1h_200mJy_closure_spectrum_average}
\end{figure*}

\begin{figure}

        \includegraphics[scale=0.6]{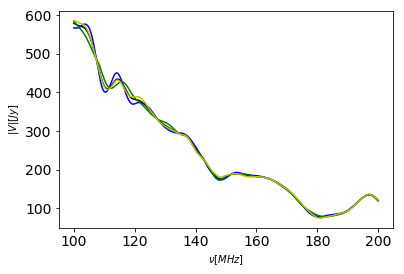}
        \includegraphics[scale=0.6]{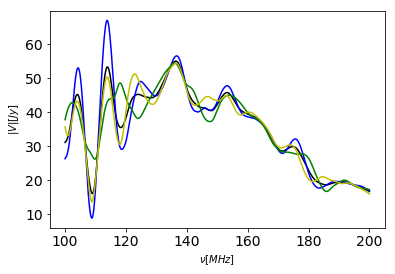}
        \includegraphics[scale=0.6]{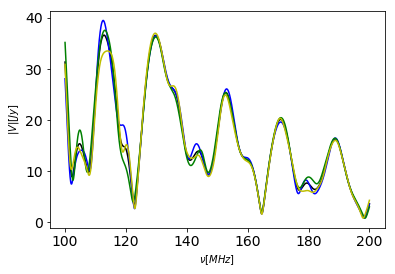}    

    \caption{Simulated visibility spectra corresponding to field~1 (top panel), field~2 (middle panel) and field~3 (bottom panel). Colours indicate baselines with different primary beams: $EE$ beams (blue), $EC$ beams (green) and $CC$ beams (yellow). Black shows the average visibility spectra.}
    \label{fig:visibilty_spectra_real_sky}
\end{figure}

\begin{figure}
 \centering
 \begin{tabular}{c}
      
 \includegraphics[scale=0.6]{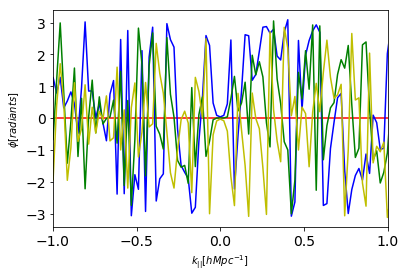}

 \end{tabular}
     \caption{Cross power spectra phases (see equation~\ref{eq:cross_power_spectra}) of simulated triads with mutual coupling for field 2 for triad $(\triangledown_{ECC},\triangledown_{EEC})$ (blue), $(\triangledown_{EEC},\triangledown_{CCC})$ (green) and $(\triangledown_{ECC},\triangledown_{CCC})$ (yellow), respectively.}
    \label{fig:cross_power_spectra}
\end{figure}

Figure~\ref{fig:full_sky_models} shows the model images of field 1, 2 and 3, together with apparent sky model obtained by applying the beam without mutual coupling $E_{H}$. Figure~\ref{fig:visibilty_spectra_components} shows visibility spectra corresponding to our sky models for the 29~m triad $\triangledown_{HHH}$. 

Fornax~A is the dominant source in field~1, and its visibility spectra are essentially the same as the case when the sky model includes both GLEAM sources and the diffuse emission (``full foreground model"). Fornax~A is, however, in the primary beam sidelobe region in field~2 and, therefore, largely attenuated, with an apparent flux density up to $\sim 8$~Jy. Field~2 is a relatively cold patch of the sky, with the Galactic plane on the far sidelobes of the primary beam. As a result, GLEAM sources in the main lobe are the dominant component, largely determining the visibility spectra of the full foreground model. Conversely, field~3 corresponds to an area of relatively bright diffuse emission, particularly at low frequencies ($\nu<120$~MHz), with the Galactic plane appearing across the second sidelobe of the beam. Beyond this range, GLEAM sources dominate, including the frequency range used for power spectra in this work, i.e. $160-190$~MHz. 

It is worth noting the striking difference between visibility spectra for field~3 in the case of the HERA ideal beam and the mutual coupling beam: although point source visibility spectra are not tremendously different, visibility spectra of the diffuse emission component oscillates significantly across the $100-200$~MHz range, with peak-to-peak variations occurring with a $\sim 5$~MHz period. The full foreground model spectrum, after attenuation by the mutual coupling beams, is far from being spectrally smooth.

Figure~\ref{fig:GLEAM_1h_200mJy_closure_spectrum_average_component} shows the closure spectra and the corresponding power spectra of each sky model component for triad $\triangledown_{CCC}$. In field~1 Fornax~A is the dominant source and appears point-like for the 29~m triad, with an approximately zero closure spectrum. Diffuse emission, however, shows closure spectra with pronounced frequency structure, likely due to emission from the Galactic plane in the beam sidelobes at $\alpha = 6-7^{\rm h}$ (see top left and right panel of Figure~\ref{fig:full_sky_models}).

This results in a $10^4-10^8$~mK$^2$~($h^{-1}$~Mpc$)^{3}$ excess power at $k_\parallel>0.5$~$h$~Mpc$^{-1}$ above the Fornax~A model. As Fornax~A is the brightest source, the power spectrum of the full foreground model closely resembles the Fornax~A one: the foreground power is largely confined at small $k_\parallel$ values, i.e. $|k_\parallel| < 0.5$~$h$~Mpc$^{-1}$, and remains flat at larger $k$ modes.

Closure spectra and power spectra for the field~2 case, are similar to field~1. GLEAM sources are the dominant foreground component at all frequencies and their closure spectra is fairly smooth in frequency. This results in a power spectrum of the full foreground model similar to field~1, with power contained at $|k_\parallel| < 0.5$~$h$~Mpc$^{-1}$. Due the Galactic plane in the sidelobe region, the closure spectrum of diffuse emission shows pronounced frequency structure, compared to the closure spectrum of GLEAM sources, resulting in a $\sim 10^8$ times higher power than GLEAM sources at $|k_\parallel| > 0.5$~$h$~Mpc$^{-1}$. 

Field~3 is a different case, where contributions from diffuse emission and GLEAM sources are at a comparable level. The closure spectrum of the full foreground model has frequency structure due to the coupling of diffuse emission and beam sidelobes. The power spectrum is different compared to the two other fields as there is foreground power up to $k_\parallel \sim 0.5$~$h$~Mpc$^{-1}$ and even beyond in the case of negative $k$ modes, with the asymmetry due to the asymmetric brightness distribution of the Galactic plane. This inevitably results in power that is between $10^2-10^7$ times higher than power from GLEAM sources at $|k_\parallel|>0.5$~$h$~Mpc$^{-1}$. With no bright source on the main lobe to ``mitigate" the leakage from diffuse emission, the composite sky model shows an excess power that can be $10^4-10^6$~mK$^2$~($h^{-1}$~Mpc$)^{3}$ at $|k_\parallel| \sim 0.5$~$h$~Mpc$^{-1}$ compared to the other two fields.  

Figure~\ref{fig:GLEAM_1h_200mJy_closure_spectrum_average} displays closure spectra and power spectra $P_\triangledown$ for the three simulated pointings for all triads. In the case of field~1, closure spectra that include mutual coupling beams, i.e. $\triangledown_{CCC}$, $\triangledown_{ECC}$ and $\triangledown_{CEE}$, have a more pronounced frequency structure compared to the  $\triangledown_{HHH}$ triad.
Power spectra of triads with mutual coupling beams show a slight broadening in the $0.1 < k_\parallel < 0.2$~$h$~Mpc$^{-1}$ compared to the triad with ideal beams, whereas all the triads have similar power spectra beyond $k_\parallel \sim  0.2$~$h$~Mpc$^{-1}$. Power spectra that include triads with mutual coupling beams have fairly similar power spectra across the whole $k_\parallel$ range, independent of the beam type. 

In the case of field~2, closure spectra from triads $\triangledown_{CEE}$, $\triangledown_{ECC}$ and $\triangledown_{CCC}$ show a more pronounced frequency structure compared to field~1, due to the presence of the Galactic plane in the sidelobe region, together with Fornax~A. Their power spectra shows excess power up to $10^4$~mK$^2$~($h^{-1}$~Mpc$)^{3}$ at $0.1 < |k_\parallel| < 0.2$~$h$~Mpc$^{-1}$ compared to the triad with ideal primary beams. Like field~1, power spectra that include triads with mutual coupling beams have fairly similar power spectra across the whole $k_\parallel$ range, independent of the beam type.

Field~3 has bright emission in the sidelobe region and, therefore, closure spectra from triads with mutual coupling beams show even more frequency structure compared with the other fields. Power spectra are brighter compared two the previous two fields for $|k_\parallel| > 0.1$~$h$~Mpc$^{-1}$, with an excess power up to $10^6$~mK$^2$~($h^{-1}$~Mpc$)^{3}$ compared to the ideal beam triad. They also tend to show some level of asymmetry between positive and negative $k_\parallel$ modes.

We also present visibility spectra of simulated triads for completeness (Figure~\ref{fig:visibilty_spectra_real_sky}) and use them to provide an estimate of the deviation from redundancy, which may have an impact on calibration. We use the absolute difference between visibility spectra of antennas affected by mutual coupling and the average visibility spectra as a metric to quantify deviations from redundancy,  averaged over the $160-190$~MHz range.
In the case of field~1, where most of the foreground emission is within the main beam, visibility spectra are fairly similar to each other and their deviation from redundancy is $\sim 2\%$. In field~2 and 3 where there are bright emissions on the sidelobe, the non-redundancy proves to be higher, with a  non-redundancy value of $\sim 6 \%$ and $\sim 7\%$ respectively, qualitatively  previous works \citep[e.g.,][]{Choudhuri2021} have also shown similar results. These values are also well within the $10\%$ non- redundancy estimated by previous studies \citep[e.g.,][]{Dillon2020}. 

 The impact that systematic effects induced by beam-to-beam variations may have on the detection of the 21 cm need a further investigation that we leave for the future. However, we looked at their effect when different triads are averaged together, like it is in actual observations \citep{Thyagarajan2020}. Rather than the power spectrum, we computed the cross-spectra $P^c_\triangledown$ between two 29~m triads with different primary beams $\triangledown$ and ${\triangledown'}$: 
\begin{equation}
    P^c_\triangledown(k_{||})= \Tilde{\Psi}_\triangledown \Tilde{\Psi}^*_{\triangledown'} \left( \frac{\lambda^2}{2 k_B} \right)^2 \left( \frac{D_c^2 \, \Delta D_c}{B_{\rm eff}} \right) \left( \frac{1}{\Omega \, B_{\rm eff}} \right).
    \label{eq:cross_power_spectra}
\end{equation}
We compute cross power spectra between triads affected by mutual coupling, namely $\triangledown_{ECC},\triangledown_{EEC}$, and the ideal triad $\triangledown_{CCC}$, and show their phase in Figure~\ref{fig:cross_power_spectra}. The cross spectrum phase shows a certain degree of incoherency across triad pairs, with variations as large as $\pi$ at the same $k$ modes. This suggests that averaging cross power spectra together may lead a suppression of systematic effects induced by mutual coupling beams - in particular considering the large number of different beams in the final HERA configuration.

\section{Discussion and conclusions}
\label{sec:conclusions}

In this work we investigate the impact that primary beams affected by mutual coupling have on closure phase, used to detect the EoR signal. We use electromagnetic simulations of HERA dishes and both a simplified and a realistic foreground model in order to perform simulations of closure spectra and its power spectra. In the simulations we specifically include antenna pairs where primary beams are different from each other. We focus only on triad separated by 29~m baselines, already used in the early analysis of HERA closure spectra \citep{Carilli2020,Thyagarajan2020}. As realistic foreground models, we include both point sources and diffuse emission. We simulated three different fields who range from a high to a low ratio between the foreground emission in the main beam lobe and in the sidelobe region. Our main conclusions may be summarized as following: 
\begin{itemize}
    \item in the presence of beams distorted by mutual coupling effects, closure spectra exhibit more pronounced frequency structure with respect to ideal beams, i.e. not affected by mutual coupling. The effect on the power spectrum of the bispectrum phase is that foreground power bleeds from small $k$ modes to intermediate modes, e.g., $0.1 < |k_\parallel| < 0.2$~$h$~Mpc$^{-1}$. Such excess power is $\sim10^3$~mK$^2$~($h^{-1}$~Mpc$)^{3}$ and does not vary significantly with different mutual coupling beam model or foreground model. Power spectra are not significantly different between models with or without mutual coupling at $|k_\parallel| > 0.2$~$h$~Mpc$^{-1}$. The presence of diffuse foreground emission that is brighter in the sidelobe region than in the main lobe exacerbates the leakage at all $k$ modes, representing a worst case scenario amongst the foreground cases simulated in this work. This result is in agreement (at least at a qualitative level) with observed ripples in closure spectra that are present when the Galactic plane appears at low elevation \citep{Carilli2020}. Wide-field, diffuse foreground emission is known to be a relevant source of power leakage outside the wedge in standard power spectrum measurementes too \citep[e.g.,][]{Thyagarajan2015,Thyagarajan2016,Kern2020};
    \item power spectra from triads that include mutual coupling beams do not significantly vary at any $k_\parallel$ mode whether they include beams that are different for different baselines or not. In other words, the main source of foreground leakage at high $k_\parallel$ modes - compared to the unperturbed beam case - is not the beam-to-beam variation for each baseline: power spectra that have essentially any combination of mutual coupling beams (including the same type for the triad) yield power spectra that are similar to each other, with a maximum difference of $\sim10^2$~mK$^2$~($h^{-1}$~Mpc$)^{3}$; 
    \item the presence of strong foreground emission in the main lobe of the primary beam helps reducing the foreground leakage at $|k_\parallel| > 0.2$~$h$~Mpc$^{-1}$ in case of mutual coupling beams, although more complete simulations which include the 21 cm signal are needed to prove that this could be an effective observing strategy;

\end{itemize}

\section*{Acknowledgements}\label{acknowledgments}
We thank an anonymous referee for useful comments on the paper. The financial assistance of the South African Radio Astronomy Observatory (SARAO) towards this research is hereby acknowledged \href{www.sarao.ac.za}{www.sarao.ac.za}. GB acknowledges support from the Ministero degli Affari  Esteri  della Cooperazione  Internazionale - Direzione Generale per la Promozione del Sistema Paese Progetto di Grande Rilevanza ZA18GR02  and the National Research Foundation of South Africa (Grant Number  113121) as part of the ISARP RADIOSKY2020 Joint Research Scheme. NK gratefully acknowledge support from the MIT Pappalardo fellowship. MGS acknowledges support from the South African Square Kilometre Array Project and National Research Foundation (Grant No. 84156).

\section*{Data availability}

The code developed for this work and the simulation output presented in the paper will be made available upon reasonable request to the corresponding author.



\bibliographystyle{mnras}
\bibliography{closure} 





\bsp	
\label{lastpage}
\end{document}